\documentstyle[twocolumn,aps,prl,epsf]{revtex} 
\input{epsf}
\begin{document}
%\bibliographysty{prsty}
%\tighten
\title{Theory for Quantum Spin and Vortex Glasses}
\author{Ferenc P\'{a}zm\'{a}ndi, Gergely T. Zim\'anyi and Richard T. Scalettar}
\address{Physics Department, University 
of California, Davis, CA 95616}
%\date{\today}
\address{\mbox{ }}
%\begin{abstract}
\address{\parbox{14cm}{\rm \mbox{ }\mbox{ }
We develop a theory for the quantum vortex glass, with {\it both} the coupling
strengths and the site energies disordered. This model is closely related to XY
spin glasses and bosons in random media. For properly chosen distributions of
the site disorder the phase diagram consists of a superfluid phase, and Weak and
Strong Glass regions, dominated by long range and local fluctuations,
respectively. The most direct experimental manifestation of the Strong Glass
transition is the lack of divergence of the corresponding non-linear
susceptibility. 
}}
\address{\mbox{ }}
\address{\parbox{14cm}{\rm \mbox{ }\mbox{ }
PACS numbers: 05.30 Jp, 67.40.Yv, 74.60.Ge, 75.10Nr}}
\address{\mbox{ }}
\maketitle

%\end{abstract}

\narrowtext

\vskip 0.7cm

A particularly active field of research these days is the quest to
understand quantum phase transitions in the presence of disorder at T=0.
Theoretical interest focussed on one dimensional transverse field Ising problems
\cite{dfisher}, on quantum critical behaviour of several spin models
\cite{bray,huse,sachdev}, and bosons in disordered media
\cite{mfisher}. 
%Non-trivial mapping procedures revealed the close connection
%between the hard core boson and spin models and the problem of vortices
%in Type $II$ superconductors. 
In one dimension scaling techniques delivered several exact results.
In higher dimensions, after the original formulation\cite{bray} 
recently quantum fluctuations were incorporated as well
\cite{huse,sachdev}. While it was possible to describe
the weakly disordered phases of the models satisfactorily, our understanding
of strongly disordered regions is still incomplete. 
Experimental studies intensified after the discovery of materials
which are credible realizations of the spin-glass models, such as the 
$LiHoYF$ compounds\cite{rosenbaum}. 
%$^{4}He$ in Aerogel,
%and lightly doped cuprates. 
In this material at the low T glass transition, presumably
dominated by quantum effects, the non-linear susceptibility $\chi_{nl}$ 
did not diverge, in contradiction to recent numerical studies\cite{huse2}.

In the present paper we develop a new technique to capture the physics of
strong disorder. In agreement with some previous approaches\cite{sachdev} 
we find {\it two types} of glass transitions, but this new method enables us
to characterize the critical behaviour even in the case of strong disorder.
One of our key results is that the {\it non-linear susceptibilty does
not diverge at criticality}. As this result is due to local fluctuations,
we expect it to hold in a wide variety of glasses, Ising as well as the XY type,
considered here.

Specifically we consider the quantum gauge glass model, which 
describes hard core bosons propagating in a magnetic field, and as such, is
believed to contain the essential physics of vortices in Type $II$
superconductors.\cite{stroud}
%From recent numerical simulations see: 
%Giamarchi and Le Doussal also considered 
Elastic vortex models in the classical limit are also argued to support 
two distinct glassy phases\cite{thierry}, with recent experiments
indicating a transition between them as the magnetic field is tuned.
%and identified a Bragg and a strongly pinned glass phase,
%again in close analogy to our picture\cite{thierry}.
%There are several recent experimental indications for a transition between 
%a weak and a strong glass as the magnetic field is tuned. 
%Measuring whether or not the corresponding non-linear susceptibility
%diverges at this transition might be a good indicator for the 
%relevance of our work.

We consider the Hamiltonian:
\begin{equation}
{\sl H} = -\sum_{i,j} J_{ij} a_{i}^{\dagger}a_{j} - \sum_{i} \mu _{i}
a_{i}^{\dagger}a_{i}
\label{H}
\end{equation}
where $J_{ij}=J_{ji}^{*}$ a complex random number for all $i<j$, 
furthermore $<J_{ij}>= J_{0}/N$, where $N$ is the number of lattice sites. and
$<|J_{ij}|^2>=J/N$. With this convention we not only have random phases along
the bonds, representing the random fluxes, but also random bond strengths.
As our results depend only on the second cumulant of the $J_{ij}$'s, the two
models are expected to give the same results. 
%Here every site is coupled 
%with every site, to facilitate an exact treatment of the problem.
$\mu_{i}=\mu+h_{i}$, where
$\mu$ controls the density of the bosons, and $h_{i}$ is a random field or
site energy, with distribution P(h) over the {\it finite} support 
$[-\Delta, \Delta]$. Finally $a_{i}$ and $a_{i}^{\dagger}$ annihilate and 
create hard core bosons, or equivalently correspond to spin-lowering and 
raising operators.

At sufficiently high values of $\mu$ there will be exactly one boson per site,
leading to the formation of a ``Mott-insulator"\cite{mfisher}, which 
is characterized by a gap above the ground state and a lack of an order
parameter. This corresponds to the paramagnet in the spin language and to 
the normal phase in the vortex case. Lowering $\mu$ below some critical 
value $\mu_{c}$ generates holes in the system. For weak disorder they 
form extended states, giving rise to a superfluid at $T=0$ (i.e. 
ferromagnetism for spins, or superfluidity for vortex systems),
whereas for larger disorder presumably a glassy phase emerges. We will test
these expectations by entering into the different ordered phases from the
high chemical potential direction. In order to map out the 
phase diagram, we first consider the behaviour of a single hole.
We then develop a many-body calculation for finite hole densities.

The Mott phase is correctly described at low energies by the single 
particle characteristics. The critical properties will be determined from
the density of states (DOS), extending our previous formalism\cite{us1}.
%The introduction of auxiliary fields allows us to express the density of states
%as a path integral for a harmonic problem. The sites are subsequently 
%decoupled by a Hubbard-Stratonovich transformation, which is exact in the
%$N \rightarrow \infty$ limit.
The DOS consists of a band 
$\rho({\tilde\omega})=-Im ~\sigma({\tilde\omega}) /\pi $
complete with the self-consistency criterion:
\begin{equation}
\sigma({\tilde\omega}) = J\langle ({\tilde\omega} + h_{i} -
J\sigma({\tilde\omega}))^{-1}\rangle_{h} ~~,
\label{sigma1}
\end{equation}
and three $\delta$ peaks at ${\tilde\omega}_{SF}$ and at 
$\pm {\tilde\omega}_{WG}$ with $1/N$ weights. These states
support superfluid and weak spin glass order.
Their location is determined respectively by: 
\begin{eqnarray}
J_{0}\sigma({\tilde\omega}_{SF})&=&J\\
\label{SF}
\langle ({\tilde\omega}_{WG} - h -
J\sigma({\tilde\omega}_{WG}))^{-2}\rangle_{h} &=& J^{-2} ~~,
\label{WG}
\end{eqnarray}
where
$\langle \cdots \rangle_{h}$ denotes averaging over the site disorder $h_{i}$.
Here ${\tilde\omega}$ is the energy measured from the center of the band.
Similar to the Bose glass problem\cite{us1}, the asymptotics of the distribution
of the site disorder plays a crucial role, because the first holes,
being bosons, fill up the sites with the deepest energy levels. 
To investigate different possibilities, we generate $P(h)$ 
%as a convolution 
%of $r+1$ independent variables with box distributions, leading to the
with the asymptotics $ \sim (\Delta-|h|)^{r}$.
As the DOS is the same for real $J_{ij}$'s, the subsequent results 
hold for XY spin glasses as well.

For $r = 0$ one finds two qualitatively different types of the DOS. 
For small values of {\it both} types of disorder ${\tilde\omega}_{SF}$ is
distinctly below the band, supporting a superfluid (SF) order. 
For strong enough bond disorder values 
%beyond the boundary given by
%$J=\Delta~({\rm cth}^{2}(\Delta/J_{0})-1)^{1/2}$,
the SF peak gets absorbed in the continuum, and the lowest energy state
becomes the ``Weak Glass" (WG) peak at ${\tilde\omega} = {\tilde\omega}_{WG}$.
In both regimes this spin glass peak remains located at the lower band edge.
For $r = 1$ and $J=0$ we also recover the Bose glass phase\cite{us1}.
For $r = 2$ a fourth behaviour is observed. 
For $\Delta/J > [27/4 \ln (4/3)]^{1/2}=1.393$ the Weak Glass state is {\it also
absorbed} into the band (the Superfluid state dissolves at
$\Delta/J_{0} = 9/4 \ln (27/16) = 1.177$).
This behaviour is present for every $r > 1$. In this case
the self-consistent equation for $\sigma$ (Eq.\ref{sigma1}) is satisfied
by the singular terms in the integral over $\int dh ~P(h)$, which are
generated by the deepest energy sites. In this ``Strong Glass" region the
exponent $\zeta$, characterizing the asymptotics of the DOS
$\rho({\tilde\omega}) \sim ({\tilde\omega} - {\tilde\omega}_{c})^{\zeta}$
changes from the well-known value\cite{wigner} of $\zeta _{WG} = 1/2$ to
$~\zeta_{SG} = r$. From the knowledge of the frequency
dependence of the DOS one can reconstruct the asymptotics of the
on-site Green's function {\it at criticality}. For the three glassy phases $g(\tau) \sim \tau^{-(\zeta+1)}$, whereas for the superfluid $g(\tau)$ decays
exponentially, as the $\omega_{SF}$ eigenvalue is separated by a gap
from the continuum. The different values of these exponents again underscore
that the transitions from the Mott Insulator into these four phases
belong to different universality classes.

\begin{figure}
\epsfxsize=3.0in
\epsfysize=2.25in
\epsffile{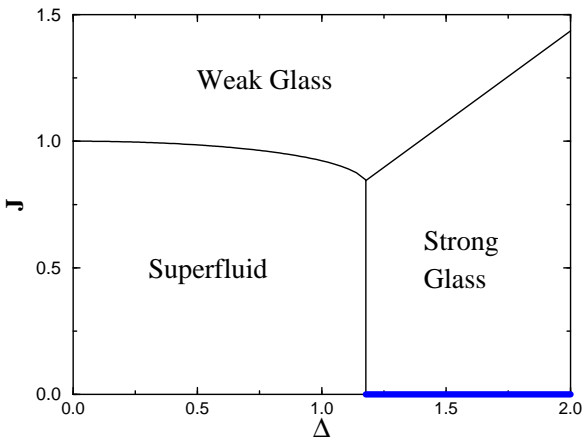}
\vskip 0.2cm
\caption{The $r=2$ phase diagram at T=0; parameters in units of $J_{0}$.
The heavy line on the $\Delta$ axis is the Bose Glass.}
\end{figure}

As the superfluid state remains separated from the glassy continuum by a gap,
just as in the clean case, it is obvious that the transition into it
stays mean field like as well\cite{us1}. Therefore in the rest of the paper
we set $J_{0}=0$ (also allowing to set $J=1$)
and concentrate on the different spin-glass transitions.
We explore the difference between these two glassy regimes by calculating
the participation ratio $P$\cite{us1} of the Weak Glass state
with the help of the cavity 
method\cite{mezard}. In this approach a single site $i$ is isolated,
then the magnetization $m_{j}$ at some other site $j$ is given as 
the magnetization of the site {\it in the absence of site $i$},
$m_{j}^{i}$, plus the effective field from $i$ times the local susceptibility
$\sigma_{j}$:
$m_{j}~=~m_{j}^{i} + \sigma_{j}~J_{ji}~m_{i} ~~.$
Using this form in the eigenvalue equation for $H$, the magnetization is 
obtained self-consistently. 
%From this the susceptibility can be recomputed, 
%leading to a self-consistency criterion. Reassuringly this equation is 
%{\it exactly the same} as the previous one for $\sigma$, whereas the
%normalization condition
%for the $m_{i}$'s coincides with the equation determining ${\tilde\omega}_{WG}$
%\cite{explanation}. 
%We introduced this method because it is more conducive to
%calculate 
We obtain for $P^{-1}=\sum_{i} m_{i}^4~$:
$~(N/P) = \langle ({\tilde\omega}_{WG} - h -
\sigma({\tilde\omega}_{WG}))^{-4}\rangle_{h}$.
The participation ratio for $r=2$, as plotted in the insert of Fig.2. clearly
exhibits an extended Weak Glass state for weak disorder, which gets localized
at the same $\Delta_{c}$ at which the Weak Glass state is absorbed into the
continuum.

\begin{figure}
\epsfxsize=3.0in
\epsfysize=2.25in
\epsffile{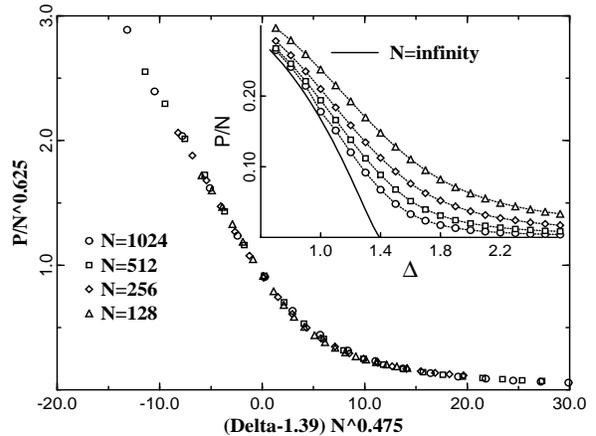}
\vskip 0.2cm
\caption{Scaling plot of the participation ratio.}
\end{figure}

The localization of the edge state of the continuum has a dramatic
effect on the critical behaviour of the non-linear susceptibility:
$\chi_{nl}^{-1} \sim 1-\langle\sigma_{i}^{2}\rangle_{h}$ \cite{suzuki}.
As seen from Eq.(\ref{WG}), for the Weak Glass state 
$1-\langle\sigma_{i}^{2}\rangle_{h} = 0$, i.e. $\chi_{nl}$ appropriately
diverges when $\mu$ approaches the edge of the band at $\omega_{WG}$ (in fact
with a $1/2$ exponent)\cite{zbyszek}. For the Strong Glass state Eq.(\ref{WG})
does not hold anymore, leading to the striking result that $\chi_{nl}$
is {\it finite} at this glass transition.

%As an independent check of our procedure we developed a numerical study
%of the participation ratio of the first hole state. 
%Since sofar the particles are treated independently,
%it suffices simply to diagonalize $H$. 
We can also compute the ground state eigenvector of $H$ numerically.
Treating the participation ratio 
as a generalized susceptibility suggests the scaling form:
$P = N^{1-\eta} f\bigl((\Delta - \Delta_{c})N^{1/{\tilde\nu}}\bigr)$, 
where $f(x)$ is a universal scaling function for a fixed $r$. 
The collapse of the curves for several system sizes in Fig.2. is a reliable
verification of this scaling assumption. In the fitting procedure we used
the theoretical value of $\Delta_{c}=[27/4 \ln (4/3)]^{1/2}=1.393$. 
The exponents were then determined as: $\eta = 0.37\pm 0.01$ 
and ${\tilde\nu}=2.10\pm 0.10$. 
%(Note that here ${\tilde\nu}$ corresponds 
%to $d\nu$ in the usual convention, i.e. its probable mean-field value is 
%${\tilde\nu =2}$). 
Both exponents describe the 
scaling of spatial correlations with the finite system size. 
As our analytic technique was conceived for $N=\infty$, presently 
the determination of their value is beyond the scope of this work. 
The one combination which {\it can} be contrasted with our calculation is
their product, appearing in the form: 
$P/N \sim (\Delta_{c}-\Delta)^{\eta{\tilde\nu}}$. The numerical procedure
gives $\eta{\tilde\nu} = 0.79$, whereas analytically we get an 
$\sim x/\ln ~x$ asymptotics, which are numerically 
indistinguishable.

So far we considered the limit of vanishing concentration of holes
in the Mott phase. Upon decreasing the chemical potential $\mu$ a finite
concentration of holes appears and a genuinely many-body treatment becomes
necessary. The usual replica trick is employed,
%to average over the randomness.
%One integrates out the bond disorder, yielding four-field couplings;
%which are subsequently decoupled by a Hubbard-Stratonovich transformation.
%For $N=\infty$ a saddle-point treatment of this action is exact, 
leading to the free energy $F=lim_{n\rightarrow 0} f_{n}/n$, where:
\begin{eqnarray}
f_{n}&=&{1\over 2}Tr~{\bf Q}^2 - {1\over N}\sum_{i}~\ln~\Phi_{i}
\label{f}\\
\Phi_{i} &=&Tr~e^{-\beta H_{i}^{0}}T_{\tau} \exp[{\bf a^{\dagger}Qa}]
\label{Phi}
\end{eqnarray}
Here the elements of the vector of annihilation operators ${\bf a}$ 
are indexed as $a_{\alpha}(\tau )$, and that of the matrix ${\bf Q}$ as $Q_{\alpha\beta}(\tau,\tau ')$.
The matrix product implies a summation over the replica indices and an
integration over the time variables. Also, 
$H_{i}^{0}=-\mu_{i}\sum_{\alpha} a_{\alpha}^{\dagger} a_{\alpha}$
and $T_{\tau}$ denotes the time ordering operator. In these equations
${\bf Q}$ satisfies the saddle-point condition:
\begin{equation}
Q_{\alpha\beta}(\tau,\tau') = {1\over N}\sum_{i}\bigl\langle 
T_{\tau} a_{\alpha}^{\dagger}(\tau) a_{\beta}(\tau ')\bigr\rangle_{\Phi_{i}}~,
\label{saddle}
\end{equation}
where ${\langle \cdots \rangle}_{\Phi_{i}}$ indicates averaging with respect
to ${\Phi_{i}}$.

Here we introduce the key technical step of our work: a coherent path integral
representation for the boson operators as follows:
\begin{equation}
e^{{\bf a^{\dagger}Qa}} = \int {D{\bf\varphi}^{*} 
D{\bf\varphi}\over Det {\bf Q}}
\exp[-{\bf \varphi^{*}Q^{-1}\varphi + \varphi a^{\dagger} + \varphi^{*} a}]
\label{coherent}
\end{equation}
With the help of these new field operators the self-consistent
equation for $Q$ transforms into:
\begin{eqnarray}
{\bf Q}(i\omega_{n})&=&{1\over N}\sum_{i}\Bigl[ {\bf E_{i}}(i\omega_{n}) - 
{\bf Q}(i\omega_{n})\Bigr]^{-1} 
\label{Q}\cr
{\bf E_{i}}(i\omega_{n}) &=& [{\bf g^{o}_{i}}(i\omega_{n}) -
{\bf \Sigma}_{i}(i\omega_{n})]^{-1} ~~~
\label{E}
\end{eqnarray} 
As the Fourier transforms of all matrices are diagonal in frequency, 
from now on these remain matrices in the replica space only. Here
${\bf [g^{o}_{i}}]^{-1}(i\omega_{n})=
(i\omega_{n}+\mu_{i})\delta_{\alpha\beta}$ is the on-site hole Green's 
function at site $i$. ${\bf \Sigma}_{i}(i\omega_{n})$ is the self energy
of the Green's function: $G_{i}(i\omega_{n})_{\alpha\beta}=
\langle \varphi^{*}_{\alpha}(i\omega_{n})\varphi_{\beta}(i\omega_{n})
\rangle_{S_{i}}$, with the local action:
\begin{equation}
S_{i} = -{\bf \varphi^{*}Q^{-1}\varphi} + \ln Tr e^{-\beta H_{i}^{0}}
T_{\tau}  e^{\bf \varphi a^{\dagger} + \varphi^{*} a} ~~.
\label{S}
\end{equation}
%This form is very hospitable for a perturbative expansion, because
%the time ordering in $G$ is contained only in the action, furthermore
%$\varphi$ being a classical field, Wick's theorem naturally applies
%for the expansion of $S_{i}$.

We expand the action $S_{i}$ up to fourth order, generating an effective
$\varphi^{4}$ theory. Using this form one can construct a perturbative
expansion for the Green's function in order to determine the self energy
${\bf \Sigma}_{i}(i\omega_{n})$. For now we stop at the level of the fully
dressed Hartree diagram, which yields a replica diagonal self-energy.
%This is so because the interaction is local in the replica space.\cite{S}
In the absence of off-diagonal elements in the self energy it is reasonable
to proceed with a replica symmetric form for ${\bf Q}$:
%\begin{equation}
$Q_{\alpha\beta}(i\omega_{n}) = \delta_{\alpha\beta}\sigma(i\omega_{n})
+\delta(i\omega_{n})\beta q $.
%\label{QQ}
%\end{equation}
Substituting this form for ${\bf Q}$ in Eq.(\ref{Q}) determines the parameters
as:
\begin{eqnarray}
\sigma(i\omega_{n})&=&{1\over N}\sum_{i}
{1\over E_{i}(i\omega_{n})-\sigma (i\omega_{n})}
\label{sigma2}\\
q&=& {q\over N}\sum_{i}{1\over (E_{i}(0)-\sigma (0))^{2}}
\label{q}
\end{eqnarray}

We summarize the solutions of these equations at $T=0$.
In the disordered phase, characterized by $q=0$, the self energy
${\bf \Sigma_{i}}(i\omega_{n})$ vanishes exponentially with T because of the
presence of the gap. In the dilute limit, after appropriate
transformations, these equations are identical with Eq.(\ref{sigma1}).
%To establish a connection with the one-particle picture
%one has to perform an analytic continuation on the $\omega$ plane and recall
%that now we measure the energies from the chemical potential.
%Then one observes that the equation for $\sigma(i\omega_{n})$
%coincides with Eq.(3), establishing the location of the phase boundary
%unequivocally.

The Weak Glass phase is easiest understood by noticing that
all formulae can in fact be expanded in the on-site disorder strength around
$\Delta=0$. In this case one finds $E=2,~~ \sigma(0)=1$ and 
$\Sigma(0) \sim q \sim (\mu_c-\mu)$.
Concerning the spectrum, at low frequencies
$E(i\omega_{n})=2+const.\times(i\omega_{n})$, reproducing the frequency 
dependence of Eq.(\ref{sigma1}); i.e. the spectrum again starts with a 
square root form
at low energies, just as at the critical point, as determined in the 
one-particle picture. Thus the Mott-to-Weak Glass transition is completely
mean-field in character. In other words, the {\it global} order parameter $q$
in Eq.(\ref{q}) is generated by comparable contributions from all sites,
indicating that the transition is driven by long range spatial correlations.

The physics of the Strong Glass transition is profoundly different.
As the chemical potential is lowered, on the right hand side ({\it rhs.}) of 
Eq.(\ref{q}), a few sites develop singular contributions.
These {\it local} instabilities have to be regulated by generating a 
non-zero value for $q$. Obviously it is impossible to expand the {\it rhs.}
around these singular points. This non-analyticity is well demonstrated
in the rather non-traditional scenario for the transition: even {\it at
the transition} the coefficient of $q$ on the {\it rhs.} of Eq.(\ref{q}) 
is less than one (since here ${\bf E_{i}=\mu_{i}}$, hence it becomes
equal to $\langle\sigma_{i}^{2}\rangle_{h}$, which is less than one outside
the Weak Glass phase). 
On the other hand, inside the glass phase (where $q>0$) the coefficient 
of $q$ must equal unity, therefore it {\it jumps} at the transition.
This profound non-analyticity signals a genuinely new type of 
transition, which is driven by local instabilities.

Next we extract the critical behaviour of $q$. There are three
types of terms to the denominator $E_{i}(0)-\sigma(0)$ in Eq.(\ref{E}). 
Sites with energies larger than $\sigma(0)$ give non-singular contributions
$\mu_{i}+ const.\times q$ just as for the Weak Glass. 
Sites with energies sufficiently below $\sigma(0)$ contribute $\sim q^{1/2}$,
whereas sites close to $\sigma(0)$ yield $\sim q^{1/3}$. 
Taking into account the width of these last two integration regions
gives $q \sim (\mu_{c}-\mu)^{4}$, with the subleading
contribution $\sim (\mu_{c}-\mu)^{6}$. 
The spectrum is computed by expanding the denominator for small frequencies.
One gets $Im \sigma \sim \omega^{1/2}$.
The spectrum is quadratic in the Mott phase and even {\it at} the transition,
i.e. the discontinuity in $q$'s coefficient in Eq. (\ref{q}) translates 
into a jump in the spectrum exponent, but the behaviour of $q$ remains 
continuous, see Fig.3. Again, the exponents of the Strong Glass criticality
depend on $r$.

\begin{figure}
\epsfxsize=3.0in
\epsfysize=2.25in
\epsffile{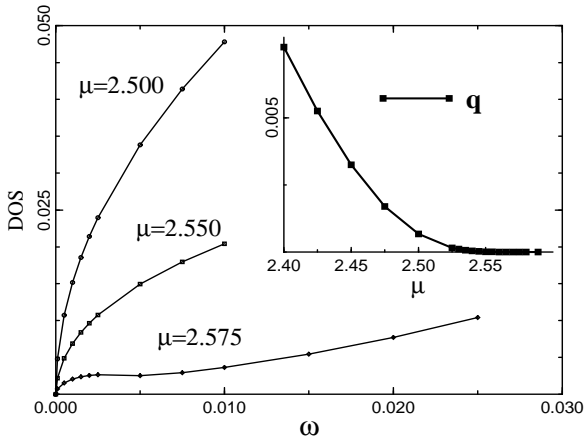}
\vskip 0.2cm
\caption{The DOS in the Strong Glass region ($\Delta=2.0$, for which $\mu_{c}=2.589$).}
\end{figure}

The following ``phase diagram" emerges from these considerations.
The transitions into the Weak and Strong Glass regions from
the Mott phase are characterized by different exponents.
The same is true approaching these boundaries from inside the glassy
phases, however these two glasses {\it do not} possess different
order parameters, hence they are not distinct phases in the thermodynamic sense.
The spin glass order parameter $q$ is non-zero in both,
and the spectrum starts with a square root form in both as well.

When crossing from the Weak to the Strong Glass, the magnitude of
$q$ is a reliable indicator, as its value drops by orders of magnitude (see
Fig.3). This is so because its asymptotic exponent of four is strongly different
from the mean field value of one. This gives rise to an anomalously
large critical region - the Strong Glass -, which is dominated by 
the local fluctuations, severely suppressing the value of $q$.
When approaching from the Mott phase, the non-linear susceptibility does 
not diverge, neither does $q$ develop an appreciable magnitude.
Here the best indicator is the low frequency spectrum.  
The Mott gap shrinks to zero at the transition. Entering the Strong Glass 
the decreasing chemical potential shifts the spectral weight further down.
To accomodate this weight at low positive frequencies, the spectrum
has to bulge up, taking a square root form: 
$\sim (\mu_{c}-\mu)\times \omega^{1/2}$, giving way
to a shifted quadratic form only at higher frequencies (Fig.3). 
Thus measuring the DOS at a low fixed frequency exhibits a gap in the 
Mott phase, and a roughly linearly increasing value inside the Strong Glass.

Lastly we comment on the issue of replica symmetry breaking.
By explicit construction we can show that the replica symmetric solution is
self-consistent in the Hartree approximation. Inspection of the structure
of the higher order diagrams reveals that {\it if} the replica symmetry 
is broken, the degree of this is at most proportional to the temperature, thus 
vanishing {\it at} $T=0$.

%The first self-energy term, which has the potential to break
%this symmetry is quadratic in the interaction of the $\varphi$ fields
%and describes the interaction of two replicas, $\alpha\ne\beta$,
%by three internal lines,
%thus being proportional to $Q_{\alpha\beta}^{3}(\omega =0)$. 
%Without this term the replica symmetric solution of the present theory 
%is self-consistent, as can be seen by explicitly constructing
%the equation for the symmetry breaking term, and demonstrate the vanishing
%of its solution. With this term the issue is open. Inspection of the structure
%of the diagrammatic expansion reveals that {\it if} the replica symmetry 
%is broken, the degree of this is proportional to the temperature, thus 
%vanishing {\it at} $T=0$.

In sum, we developed a theory for the quantum vortex glass, closely related to
XY magnets and bosons in disodered media. We included disorder {\it both}
for the coupling strengths and for the site energies. 
Utilizing one particle, many body and numerical methods we determined the 
phase diagram and the critical behaviour. We found a superfluid
phase and Weak and Strong Glass regions, the physics of which
is dominated by long range and local fluctuations, respectively. 
It was possible to go beyond previous approaches
in the strongly disordered regime by introducing a coherent field 
representation for the hard core bosons. The most accessible experimental
manifestation of the Strong Glass transition is the lack of divergence in the
corresponding non-linear susceptibility. Analogous experimental findings have
been reported in quantum spin-glass and vortex systems.

We acknowledge useful discussions with S. Sachdev.
This work has been supported by NSF-DMR-95-28535.


\vskip -0.5cm
\begin{references}


\bibitem{dfisher} D.S. Fisher, Phys. Rev. Lett. {\bf 69}, 534 (1993).

\bibitem{bray} A.J. Bray and M.A. Moore, J. Phys. C {\bf 13}, L655 (1980);
Y.Y. Goldschmidt and P.Y. Lai, Phys. Rev. Lett. 
{\bf 64}, 2567 (1990).

\bibitem{huse} J. Miller and D. Huse, Phys. Rev. Lett. {\bf 70}, 3147 (1993).

\bibitem{sachdev} J. Ye, S. Sachdev and N. Read, Phys. Rev. Lett. {\bf 70}, 4011
(1993); Phys. Rev. B. {\bf 52}, 384 (1995).

\bibitem{mfisher}
M.P.A. Fisher, P.B. Weichman, G. Grinstein, and D.S. Fisher, 
Phys. Rev. B {\bf 40}, 546 (1989).

%\bibitem{fisherlee}  M.P.A. Fisher and D.H. Lee, Phys. Rev. B {\bf 39},
%2756 (1989)

\bibitem{rosenbaum}
W. Wu, D. Bitko, T.F. Rosenbaum and G. Aeppli, Phys. Rev. Lett. {\bf 71},
1919 (1993)

%\bibitem{reppy} J.D. Reppy, J. Low Temp. Phys. {\bf 87}, 205 (1992).

%\bibitem{keimer} B. Keimer {\it et al.}, Phys. Rev. Lett. {\bf 67}, 
%1930 (1991).

\bibitem{huse2} M. Guo, R.N. Bhatt and D.A. Huse, Phys. Rev. Lett. {\bf 72},
4137 (1994); H. Rieger and A.P. Young, {\it ibid.} {\bf 72}, 4141 (1994).

%\bibitem{young} J.D. Reger, T.A. Tokuyasu, A.P. Young and M.P.A. Fisher,
%Phys. Rev. B {\bf 44}, 7147 (1991).

\bibitem{stroud} C. Ebner and D. Stroud, Phys. Rev. B {\bf 31}, 165 (1985).

\bibitem{thierry} T. Giamarchi and P. Le Doussal, Phys. Rev. Lett. {\bf 72},
1530 (1994).

\bibitem{us1} F. P\'azm\'andi, G.T. Zimanyi and R.T. Scalettar, Phys. Rev. Lett.
{\bf 75}, 1356 (1995).

\bibitem{wigner} E. Wigner Ann. Math. {\bf 65}, 203 (1957).

\bibitem{mezard} M. Mezard, G. Parisi and M. Virasoro, {\it Spin glass 
theory and beyond}, World Scientific, Singapore, p.65 (1987).

\bibitem{suzuki} T. Yamamoto and H. Ishii, J. Phys. C {\bf 20}, 6053 (1987).

\bibitem{zbyszek} F. P\'azm\'andi and Z. Doma\'nski, Phys. Rev. Lett. 
{\bf 74}, 2363 (1995).

\end{references}
\end{document}